# A Model for Personalized Keyword Extraction from Web Pages using Segmentation

K.S.Kuppusamy
Department of Computer Science
School of Engineering and Technology
Pondicherry University
Pondicherry
India

G.Aghila
Department of Computer Science
School of Engineering and Technology
Pondicherry University
Pondicherry
India

## ABSTRACT
The World Wide Web caters to the needs of billions of users in heterogeneous groups. Each user accessing the World Wide Web might have his / her own specific interest and would expect the web to respond to the specific requirements. The process of making the web to react in a customized manner is achieved through personalization. This paper proposes a novel model for extracting keywords from a web page with personalization being incorporated into it. The keyword extraction problem is approached with the help of web page segmentation which facilitates in making the problem simpler and solving it effectively. The proposed model is implemented as a prototype and the experiments conducted on it empirically validate the model's efficiency.

## Keywords
Keyword extraction, web page segmentation, web personalization.

## 1. INTRODUCTION
The World Wide Web houses mammoth collection of information which can be harnessed by people with varying informational requirements. The process of fetching the information from this gargantuan sized resource is made easier by the web search engines.

A user surfing the World Wide Web might expect the web to respond to his /her personal information requirement context. The first step towards achieving this goal is the process of incorporating personalization in the surfing procedure. The personalized web surfing can be realized by building user profiles. The user profiles can be built either explicitly generated by the system in an implicit manner.

This paper proposes an approach to extract user specific keywords from web pages. The keywords extracted using this procedure serves as an extended tool to understand the user's informational requirement context. The approach proposed in this paper includes the weight assignment to the keywords so that ordering can be introduced and more relevant keywords would be made to top the list which may be further used in recognizing the user preferences.

The keyword extraction procedure illustrated in this paper harness the web page segmentation technique. Due to the incorporation of segmentation in to the keyword extraction the structural semantics of the web page are harnessed.

The objectives of this research paper are as listed below:

- Proposing a model for personalized keyword extract from web pages.
- Incorporating segmentation in to the process to harness the structural semantics of the web page.

The remainder of this paper is as organized as follows: Section 2 explores various motivational works which are carried out in this domain. Section 3 illustrates the mathematical model and the algorithms. In section 4 experiments and result analysis are discussed. Section 5 lists out the conclusions and future directions for this research work.

## 2. MOTIVATIONS
This section explores various motivational works carried out in this domain, which forms the foundations for the model proposed in this research work.

As with any other area of study in the World Wide Web, the keyword extraction from the web pages is also an active area of research in the information retrieval domain. The proposed model incorporates two sub-domains of studies. They are as listed below:

- Keyword Extraction
- Web Page Segmentation

### 2.1 Keyword Extraction
There exist many research works on extracting keywords from web pages. This section explores few of the motivational works carried out in this domain.

The keyword extraction is approached using statistical measures of word count etc [1]. A word co-occurrence based approach is illustrated in [2]. Methods based on the Artificial Intelligence techniques were proposed by [3],[4].

There exist recent research works in keyword extraction which utilizes the World Wide Web and search engines to identify the relationship among the words [5], [6]. These models utilize the result count of search engines to identify important keywords.

### 2.2 Web Page Segmentation
Web page segmentation is an active research topic in the information retrieval domain in which a wide range of experiments are conducted. Web page segmentation is the process of dividing a web page into smaller units based on various criteria. The following are four basic types of web page segmentation methods. They are

- Fixed length page segmentation
- DOM based page segmentation
- Vision based page segmentation
- Combined / Hybrid method

A comparative study among all these four types of segmentation is illustrated in [7]. Each of above mentioned segmentation methods have been studied in detail in the literature. Fixed length page segmentation is simple and less complex in terms of implementation but the major problem with this approach is that it doesn't consider any semantics of the page while segmenting. In DOM base page segmentation, the HTML tag tree's Document Object Model would be used while segmenting. An arbitrary passages based approach is given in [8]. Vision based page segmentation (VIPS) is in parallel lines with the way, humans views a page. VIPS [9] is a popular segmentation algorithm which segments a page based on various visual features.

Apart from the above mentioned segmentation methods a few novel approaches have been evolved during the last few years. An image processing based segmentation approach is illustrated in [10]. The segmentation process based text density of the contents is explained in [11]. The graph theory based approach to segmentation is presented in [12].

## 3. THE MODEL
This section explores the proposed personalized model for keyword extraction from web pages. The proposed model is as depicted in Fig 1. The model has following components:

- The Segmentor component is responsible for splitting the page into various segments. A variation of the DOM based segmentation is followed in the proposed model.
- The content analyzer receives each of the segments. The text contents of the segments are stripped from their html tags. These text contents are then analyzed through Yahoo! Content Analysis API (YCA) .[13]
- The Segment Scorer component calculates the score of each segments against the words specified in the profile bag. The scoring procedure is a variation of MUSEUM (Multi Dimensional Segment Evaluation Method) proposed by us earlier.[14]
- The keyword scorer component receives inputs from both the content analyzer and segment scorer components. The keyword scorer computes the new weight by fusing the scoring generated by content analyzer and segment scorer.

### 3.1 The Mathematical Model
This section explores the mathematical representation of the proposed model for personalized keyword extraction from web pages.

The source of the page is denoted as $\Omega$. The source page is split in to various segments as shown in (1).

$$\Omega = \{\omega_1, \omega_2, \omega_3 ... \omega_n\} \quad (1)$$

In (1) each $\omega_i$ represents a segment of the web page. The text contents of $\omega_i$ are separated from the html tags as shown in (2).

$$\Omega = \{\forall_{i=1..n} : \omega_i = \Gamma(\omega_i)\} \quad (2)$$

In (2) $\Gamma(\omega_i)$ represents the function to strip textual contents from the html tags. This step is performed to make the content analyzer to consider only the textual contents and omit the tags which are used for formatting the contents.

After the removal of tags, the contents are submitted to content analysis service. The content analysis service returns an array which holds both the significant terms and their weight, as shown in (3).

$$\Psi = \left\{ \bigcup_{i=1}^{n} \lambda(\omega_i) \right\} \quad (3)$$

In (3), $\lambda(\omega_i)$ denotes the Yahoo! Content Analysis Service call. The array $\Psi$ would hold the output of content analyzer. Each item in the output is a pair as shown in (4).

$$\Psi = \{\forall_{i=1..n} \lambda(\omega_i) : \langle \alpha_1, \delta_1 \rangle, \langle \alpha_2, \delta_2 \rangle ... \langle \alpha_n, \delta_n \rangle\} \quad (4)$$

Each $\langle \alpha_i, \delta_i \rangle$ pair of significant term and its weight.

The user's profile bag is represented with a set of keywords as shown in (5).

$$\Upsilon = \{\beta_1, \beta_2 ... \beta_n\} \quad (5)$$

The segments of the page weighed against these profile keywords $\beta_i$. The segment scorer component of the proposed model calculates the weights of the segments using a multi-dimensional approach.

The various dimensions with which the segments are evaluated with the profile keywords are as shown in (6).

$$\Theta = \{L, I, V, T\} \quad (6)$$

In (6) L indicates Link, I indicates Image, V indicates Visual Weight and T indicates Theme weight.

Each segment $\omega_i$ is evaluated for its significance along all the four dimensions specified in (6).

$$\Lambda = \left\{ \bigcup_{i=1}^{n} [L(\omega_i) + I(\omega_i) + V(\omega_i) + T(\omega_i)] \right\} \quad (7)$$

In (7), $\Lambda$ indicates the segment score array. As depicted in (7) weight of each segment is calculated by adding the individual dimensions of $\Theta$.

These segment weights need to be compared with the weighted significant terms to extract the personalized keywords from the list.

$$\Psi_P = \begin{Bmatrix} \omega_1[\langle\alpha_1,\delta_1\rangle,\langle\alpha_2,\delta_2\rangle...\langle\alpha_n,\delta_n\rangle] \\ \omega_2[\langle\alpha_1,\delta_1\rangle,\langle\alpha_2,\delta_2\rangle...\langle\alpha_n,\delta_n\rangle] \\ ... \\ \omega_n[\langle\alpha_1,\delta_1\rangle,\langle\alpha_2,\delta_2\rangle...\langle\alpha_n,\delta_n\rangle] \end{Bmatrix} \oplus \begin{Bmatrix} \Lambda(\omega_1) \\ \Lambda(\omega_2) \\ ... \\ \Lambda(\omega_n) \end{Bmatrix} \quad (8)$$

is more than the threshold value then it is added in to the keyword repository, else they are ignored as shown in (9).

$$\Psi'_P = \Psi'_P \cup \{\forall_{i=1..n} \Psi_P(i) > \rho\} \quad (9).$$

The corresponding terms in $\Psi_P(i)$ whose weight values are greater than the threshold $\rho$ are included in the final list of keywords.

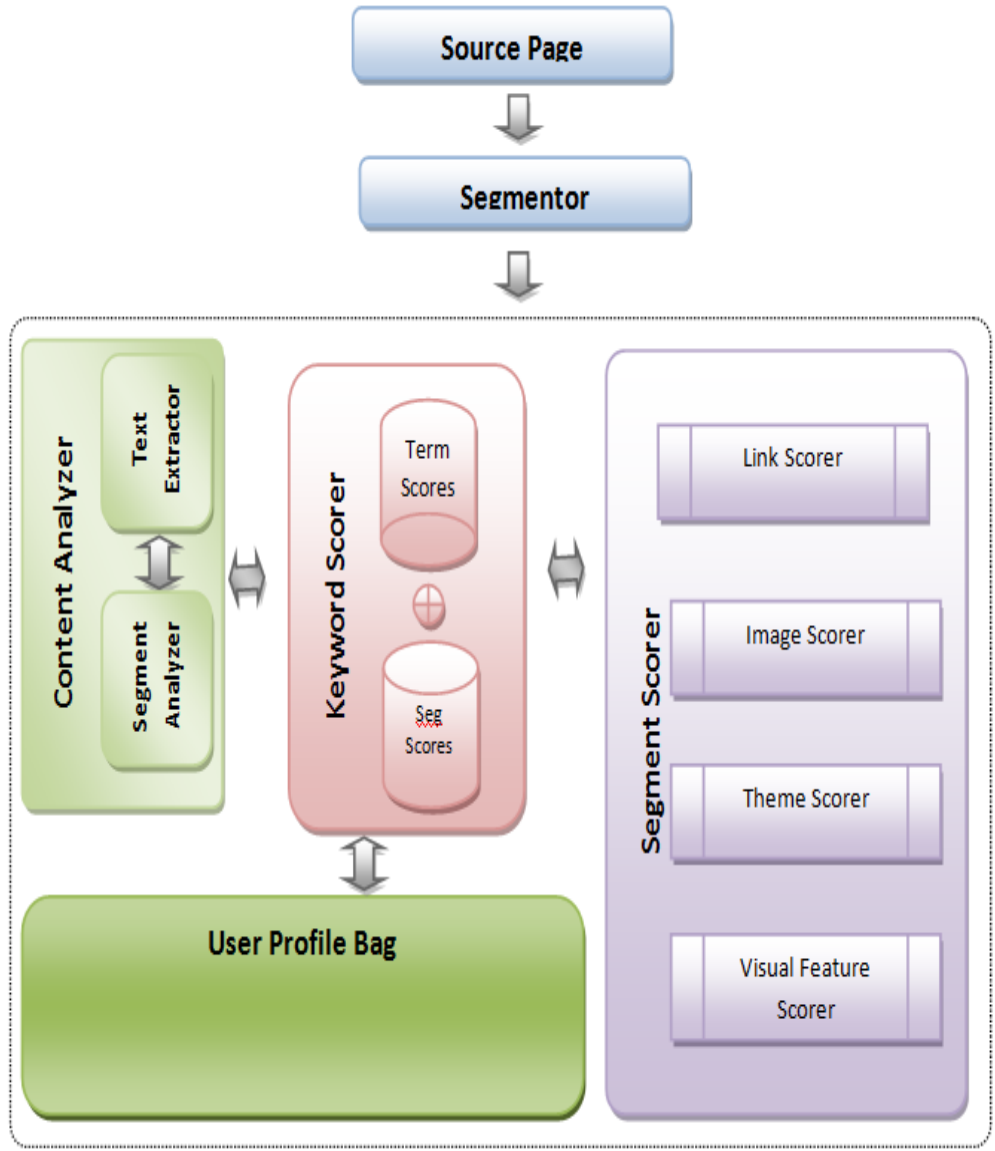

**Fig 1: Block Diagram of the Proposed Model**

In (8), each $\omega_i[\langle\alpha_1,\delta_1\rangle,\langle\alpha_2,\delta_2\rangle...\langle\alpha_n,\delta_n\rangle]$ represent the weighted significant terms extracted for the segment $\omega_i$. The personalized weights for each segment is denoted by $\Lambda(\omega_1)$. The symbol $\oplus$ represent the score fusion operator.

The weights calculated by both the segments are summed up to calculate the personalized weight. If the calculated weight

### 3.2 The Algorithm
The algorithmic representation of the above said process is illustrated in this section.

Algorithm PersoanlizedKeywords

Input: User Profile, Source Page

Output: Personalized Keyword Array

Begin

1. Segment the source page into segments.
2. for each segment i

    call t [] = extractText(segment i);

    call k [] = yca(t);

    call l = linkWeight(segment i)

    call im = imageWeight(segment i)

    call vf = visual weight(segment i)

    call th = themeWeight(segment i)

    segWeight [] = l + im + vf + th;

3. for each element in k

    merge the yca score and segWeight

    mScore[]=yca[]+segWeight

4. initialize the personalized keyword array pk = null
5. for each element in mScore

    if mScore[] > threshold then

      pk[] = pk[] . key(mScore)

6. return pk

End

In the algorithm, the extractText() function removes the html tags from the segment and returns the simple text. The yca() receives the text as input and returns the significant terms with their scores.

**Table 1. Experimental Results**

| User Group ID | Mean Initial Profile terms Count | Mean Segment Count | Mean Personalized Keywords Count |
|---|---|---|---|
| 1 | 4.32 | 8.12 | 9.43 |
| 2 | 5.43 | 12.14 | 5.65 |
| 3 | 7.45 | 14.13 | 8.32 |
| 4 | 4.38 | 17.12 | 6.32 |
| 5 | 8.12 | 14.38 | 8.32 |
| 6 | 4.13 | 14.65 | 6.45 |
| 7 | 7.12 | 16.72 | 9.12 |
| 8 | 8.11 | 15.43 | 10.25 |
| 9 | 9.10 | 14.25 | 11.14 |
| 10 | 5.65 | 16.14 | 8.24 |
| 11 | 4.34 | 17.18 | 6.75 |
| 12 | 5.32 | 15.14 | 7.38 |
| 13 | 6.12 | 14.85 | 8.32 |
| 14 | 7.11 | 9.15 | 10.15 |
| 15 | 6.12 | 10.25 | 8.34 |

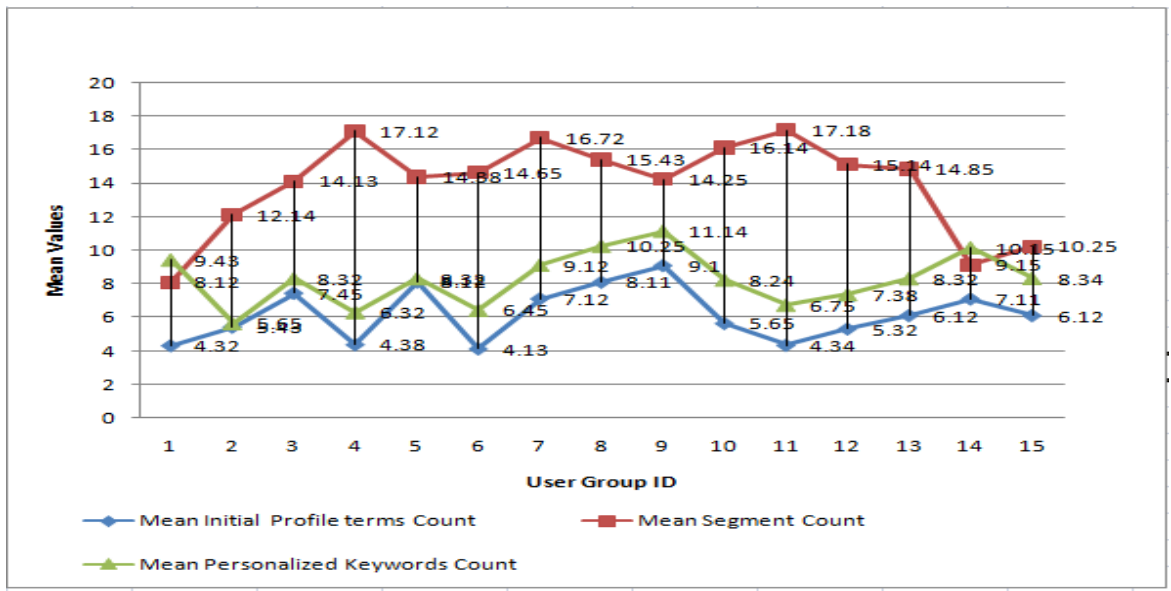

**Fig 2: Initial Profile terms Vs Personalized Terms**

## 4. EXPERIMENTS AND RESULT ANALYSIS

The proposed model is implemented as a prototype to experimentally validate the model. The prototype implementation is done with the software stack including Ubuntu Linux, Apache, MySql and PHP. For client side scripting JavaScript is used. With respect to the hardware, a Core i5 processor system with 3 GHz of speed, 8 GB of RAM is used. The internet connection used in the experimental setup is a 128 Mbps leased line.

The results of the experiments are tabulated in Table 1. The data in Table 1 illustrates data for groups of users. The second column of the table is mean initial profile terms count which is the average number of profile terms initially there for this group of users.

The third column represents the mean segment count which indicates the average number of segments of the page for those groups of users.

The last column of the table lists out the mean personalized keywords count which represent mean of the number of keywords given as output for this user group.

The experimental results which are listed out in Table 1, considers user as group. This approach is followed so that the data would cover more number of users rather than recording data for small number of users.

For the current experimentation purpose each group consisted of ten users. So the overall result data covers data for 150 users whose informational requirements would be divergent.

The chart in Fig.2 confirms the fact that higher the initial number of profile terms more the personalized keywords extracted from the page.

The mean of initial profile terms count of across the group in our experiments in 6.19. The mean of the personalized terms extracted from the page across the group is 8.28 which indicate that the profile terms extracted outnumbers the initial terms supplied by the user.

For users with different profile terms, the keywords extracted also differ, as the segment evaluation is based on profile terms.

## 5. CONCLUSIONS AND FUTURE DIRECTIONS

This section lists out the conclusions and future directions of this research work. The conclusions derived out are as listed below:

- The proposed segmentation based keyword extraction model, exploits the existing user terms, segmentation techniques to build the final list of personalized keywords.

- The proposed model can be used to better represent the user's information requirement context through the use of personalized keywords extracted.

The future directions for this research work are as listed below:

- The proposed model can be used with ontology based user profile representation techniques like FOAF to further improve the efficiency.

- The proposed model can be further enriched by the incorporation of Natural Language Processing techniques.

## AUTHORS PROFILE

**K.S.Kuppusamy** is an Assistant Professor at Department of Computer Science, School of Engineering and Technology, Pondicherry University, Pondicherry, India. He has obtained his Masters degree in Computer Science and Information **Technology** from Madurai Kamaraj University. He is currently pursuing his Ph.D in the field of Intelligent Information Management. His research interest includes Web Search Engines, Semantic Web. He has made 8 international publications.

**G. Aghila** is a Professor at Department of Computer Science, School of Engineering and Technology, Pondicherry University, Pondicherry, India. She has got a total of 22 years of teaching experience. She has received her M.E (Computer Science and Engineering) and Ph.D. from Anna University, Chennai, India. She has published more than 55 research papers in web crawlers, ontology based information retrieval. She is currently a supervisor guiding 8 Ph.D. scholars. She was in receipt of Schrneiger award. She is an expert in ontology development. Her area of interest includes Intelligent Information Management, artificial intelligence, text mining and semantic web technologies.